# Building Bridges – AI Custom Chatbots as Mediators between Mathematics and Physics

Julia Lademann[1,*] (ORCID ID 0009-0009-3033-4021), Jannik Henze[2], and Sebastian Becker-Genschow[1]

[1]*University of Cologne, Faculty of Mathematics and Natural Sciences, Digital Education Research, 50931 Cologne, Germany*

[2]*University of Cologne, Faculty of Mathematics and Natural Sciences, Institute for Physics Education, 50931 Cologne, Germany*[1]

**ABSTRACT.** This work explores the integration of AI custom chatbots in educational settings, with a particular focus on their applicability in the context of mathematics and physics. In view of the increasing deployment of AI tools such as ChatGPT in educational contexts, the present study examines their potential as personalized tutoring systems. The study assesses the impact of AI-generated learning materials on the learning experiences and performance of sixth-grade students, with a particular focus on proportional relationships in mathematical and physical contexts. The randomized controlled study with $N = 214$ students compared traditional textbook materials with explanations generated by a custom chatbot. The results demonstrated that while AI-generated materials had an indefinite impact on learning outcomes, they significantly enhanced positive-activating emotions, situational interest, and self-efficacy, while reducing intrinsic and extrinsic cognitive load. These findings underscore the potential of AI to transform educational practices by fostering a superior learning experience. However, further research is required to clarify its impact on learning performance and long-term learning outcomes. The study highlights the importance of careful integration and customization of AI tools to maximize their benefits in physics education.

## I. INTRODUCTION

Since the release of ChatGPT in 2022, there has been growing attention to how AI chatbots can be used profitably in an educational context, and the integration of artificial intelligence (AI) is emerging as a key factor at both the school and university level [1]. For instance, AI tools can be utilized by educators to develop lesson plans, create differentiated learning materials, and provide feedback. Additionally, they can be employed by students as intelligent tutoring systems, research or writing tools [2–5]. Especially Large Language Models (LLMs) have the potential to transform educational experiences, particularly in specialized fields such as physics [1]. LLMs are a category of artificial intelligence based on neural networks, trained on vast datasets. Their objective is to comprehend and generate text in a manner that imitates human communication, thereby opening up a multitude of opportunities for educational applications [3]. By enabling more personalized and interactive learning approaches, these models have the potential to transform the way students learn. One of the most widely recognized applications of LLMs is in the field of chatbots, which are software tools that engage with the user in written dialogue [1]. In STEM education, chatbots can be utilized as learning assistants to provide feedback to e.g. exercises and individual topic related questions. They can generate a variety of additional exercises which also can be adapted to the individual students learning status. In natural sciences, chatbots can assist students during experiments [2,6–9]. However, despite this promise, the question of how exactly learning with AI chatbots affects students' learning performance and experience remains unanswered [5]. This indicates the necessity for further research in this field.

Moreover, regarding the use of AI chatbots in school, challenges, including a lack of in-depth contextual understanding, the difficulty in assessing the quality of responses and a deficiency in higher-order cognitive abilities are mentioned [10]. For instance, learners frequently utilize chatbots without reflection when engaged in physics tasks, without attempting to solve the tasks independently [11]. These findings highlight the necessity for an in-depth analysis of responses generated by ChatGPT, particularly with regard to their scientific veracity, by both instructors and students [12] and underscore the need for cautious incorporation of AI into the education system by conducting a thorough and comprehensive evaluation of this potentially profitable tool.

The present study is therefore focused on exploring the use of AI chatbots that provide individual feedback on questions or solutions from students. The aim of this study is to conduct a comparative analysis of traditional and AI-generated learning material, examining the impact of AI-generated explanations on students' learning experience and performance in physics.

---

[*]julia.lademann@uni-koeln.de

The primary findings of the study indicate that the use of AI-generated material has, while no discernible positive or negative effects on learning effectiveness could be identified, a notable impact on the learning experience of students. Emotionally, in the experimental group higher levels of positive-activating emotions were measured. Moreover, self-efficacy expectations and situational interest were also significantly higher in the experimental group compared to the control group. Conversely, the levels of intrinsic and extrinsic cognitive load were significantly lower.

## II. STATE OF RESEARCH

### A. Context of physics teaching

Students' comprehension problems in physics lessons are often related to a lack of mathematical skills [13–15]. However, these difficulties are also frequently attributed to the challenge of transferring the skills acquired in mathematics to the physical context [16,17]. The causes of these difficulties are not uniform across all students – they may arise from misconceptions about the physical meaning of, for example, ratios, products, functions and neutral elements [18].

Moreover, in the fields of physics and mathematics, comprehension problems potentially emerge from difficulties in interpreting and linking different forms of representation, which can be defined as multiple representations [19]. These forms of representation, including formulas, graphs, diagrams, and tables, are a means of encoding important information. It is therefore beneficial for learners to possess the capacity to interpret and transition between the various forms, thereby facilitating a full comprehension of scientific concepts [19–21]. By providing targeted assistance in establishing connections between the structures of depicted forms of representation, students are more likely to develop an understanding of the concepts and are less likely to hold misconceptions [20].

As an example also representing the problem of transferring mathematical knowledge into physics, the topic "proportional relationships" was chosen for this study. It is taught in seventh-grade mathematics in Germany. A proportional relationship has various forms of representation (graph, table, formula) and students must develop the ability to switch between these representations without difficulties. To do so, students have to interpret the mathematical models from a physics point of view or to translate physical behavior in a mathematical context [22,23]. Hence, the transfer of mathematical knowledge is necessary to interpret the different forms of representation and switch between them. AI chatbots with vision capabilities like ChatGPT may offer a potential solution to this issue [2,3,5] as they are able to read and analyze images of representations. In this manner, learners can be aided in both the interpretation of discrete forms of representation and in the recognition of the encoded information and the relationships between different forms of representation.

### B. AI Chatbots in Education

*1. Learning Experience and Learning Performance*

The integration of AI into the field of education is rapidly transforming teaching and learning experiences. AI is gaining recognition for its ability to improve educational outcomes, thereby fostering a growing interest in research within the domain of AI in education in recent years [24,25]. Generative AI in particular opens up new opportunities for the design of teaching and learning environments through the personalized assistance of chatbots. AI chatbots can provide learners with individualized feedback and address specific learning difficulties and questions [2,3,5]. They have the potential to enhance student productivity and foster motivation [4,26]. Moreover, a meta-analysis demonstrates that AI chatbots in general can significantly impact student learning outcomes [27] and result in advantages such as better access to information and a simplification of personalized and complex learning [10]. ChatGPT for instance, is not only capable of solving tasks, but can also explain solutions and approaches and create tasks itself [7,8]. This additionally gives learners the opportunity to generate a variety of exercises and to comprehend solution paths. Moreover, personal chatbots can be configured to operate as so-called custom chatbots, thereby providing users with assistance and feedback. However, particularly within the school context, the potential of AI chatbots, especially custom chatbots, as intelligent tutoring assistants is still largely underexplored. One remaining question is how learners perceive AI chatbot-generated feedback and their interaction with AI chatbots, and whether it influences the learning process [5]. An increasing number of studies are investigating the influence of the use of AI chatbots, in particular ChatGPT, on the learning experience and learning performance.

It was found that the utilization of ChatGPT fosters a low-pressure environment, thereby encouraging learners to seek further clarification and assistance with greater ease [28]. Learners exhibited greater satisfaction due to the assurance of privacy. Additionally, learning with chatbots may result in the cultivation of positive emotions and an increased sense of well-being [29]. In the field of mathematics, the utilization of ChatGPT has been demonstrated to enhance self-efficacy and to improve the development of conceptual understanding [30]. Similarly, in the field of physics, ChatGPT has been shown to positively influence both learning experiences and learning performance [31,32]. Moreover, it was found that employing ChatGPT for experimentation in school helped to correct misconceptions in physics and facilitated a more profound understanding of physical concepts [9]. Concerning research-methods, it was shown by a comparative analysis between the utilization of ChatGPT and conventional search engines that LLMs can reduce the mental effort demanded by learners in research tasks, yet simultaneously result in a decline in the quality of reasoning and conclusions [33].

At this point, it is unclear to what extent learners are able to identify errors in AI-generated content and whether learning with ChatGPT fosters critical thinking or, conversely, reinforces misconceptions [5]. The persuasive manner in which ChatGPT presents information can obscure inaccuracies. Therefore, learners should not rely on it as their sole source of information [34]. Nevertheless, nearly half of the students enrolled in an undergraduate-level introductory physics course expressed confidence in the ability of ChatGPT to provide accurate responses, regardless of their accuracy [35]. Conversely, the utilization of an AI assistant in the context of experimentation presents an effective method for the rectification and resolution of potential misconceptions within the field of physics education, as well as facilitating a more profound comprehension of fundamental physics concepts [9].

In conclusion, AI chatbots such as ChatGPT are valuable tools, but their integration into physics teaching must be handled with care. They offer the advantage of enabling personalized learning and reducing the workload of teachers. However, despite these advantages, further research is needed to identify successful didactic concepts and meaningful implementations in physics teaching [1].

### *2. The capabilities of ChatGPT in physics*

When discussing the potential use of AI chatbots in an educational setting, the question of the accuracy of AI-generated responses cannot be overlooked. These should provide accurate answers to technical queries with a high degree of probability. This is particularly crucial in mathematical and physical contexts, as errors can be easily overlooked and are challenging for students to verify. Various studies are currently examining the capabilities of AI chatbots in a physical context.

For instance, ChatGPT-4 demonstrated high accuracy in solving the FCI as well as concept tasks in the field of mechanics. Its performance exceeded that of engineering students [36]. Even ChatGPT-3, which was able to pass an introductory physics course, only made mistakes that resemble those of beginners [37]. Subsequently, it was demonstrated that ChatGPT-4 has achieved a significantly higher score in the aforementioned context. Indeed, the responses exhibit a level of competence that is nearly indistinguishable from that of an expert, with a few notable exceptions and limitations [38]. The studies demonstrate that ChatGPT-4 has proficient fundamental physical capabilities. However, they also indicate that there is an upward trend in performance across the models. Nevertheless, a more profound comprehension is occasionally absent, particularly in the capacity to prove theorems or derive physical laws [34].

As a vision-capable chatbot, ChatGPT-4 is also able to analyze images. This can be particularly interesting for mathematics and physics, since graphs, tables, and formulas are often difficult to describe or to transfer into the chatbot's text field. However, a study examining ChatGPT-4's capacity to solve the TUK test revealed that while ChatGPT-4 typically accurately describes the approaches to solving graph-related tasks, it exhibits deficiencies in graph analysis. For instance, it fails to discern intersections with the axes correctly [39]. However, a trend towards improvement could be seen here as well, since ChatGPT-4o was able to achieve significantly better results than its predecessor ChatGPT-4 and also outperforms all other tested models [40]. This observation also shows a successive improvement in LLMs for the physical learning context.

### *3. Custom Chatbots*

Despite the existence of a multitude of chatbots, it is often necessary to adapt them to achieve the desired performance in a given application [36]. A study with electrical engineering students showed that learning with a custom AI chatbot (based on ChatGPT) significantly improved learning performance and self-efficacy, both in comparison to traditional learning methods and to learning with ChatGPT. Students reported heightened confidence and effectiveness in utilizing the custom chatbot as a learning assistant [41]. This motivates a systematic configuration and evaluation of chatbots for implementation in educational settings.

For the educational context, a good option to tailor a chatbot to serve a specific purpose is called Augmentation. This method does not require detailed knowledge in machine learning and neural networks [42]. The chatbots are based on already existing LLMs and adapted to a specific purpose. A way of augmentation is called Retrieval Augmented Generation (RAG) which is a method used to enhance and specify the capabilities of LLMs by integrating external data sources [42,43]. Documents that serve as the external knowledge base are divided into smaller segments and converted into an embedding. When a user submits a prompt, it is compared with the document embeddings. Selected segments are combined with the user's prompt and sent to the LLM as a new prompt reading: "Reply to [user prompt] using the following background materials: [relevant text segments]." [42]. The LLM then generates a response based on both the user prompt and the external knowledge. RAG provides a practical method for customization for specific use cases by curating the database with relevant materials. The reliance on external documents reduces the risk of hallucinations [42]. Therefore, in educational contexts, RAG can be employed to create customized chatbots that provide precise answers based on course-specific materials, such as lecture notes, syllabi, or problem sets. By setting up a separate RAG instance for each course, institutions can ensure that the chatbot responds accurately and consistently within the intended domain [42]. A variety of platforms, for example operated by OpenAI[2] and Anthropic[3], offer the possibility of augmentation based on their respective LLM, thereby enabling the creation of custom chatbots aligned with specific requirements and domains of use. This is a relatively straightforward process for teachers or students to configure their own custom chatbots. The custom chatbots at OpenAI are designated as GPTs. The users may instruct their GPT to behave and react in a desired manner, such as using situation-specific language or a certain length of answers, by means of a configuration prompt that they define themselves. Moreover, subject- or situation-specific knowledge can be provided to the GPT by uploading corresponding files. For this study we decided to use the method of Augmentation and OpenAI to customize an own GPT (see **IV.A.3.**).

## III. RESEARCH QUESTIONS

The current state of research implies that further studies in this still new and under-researched area are both useful and imperative. AI chatbots have the potential to address learning deficiencies among students and facilitate personalized learning. However, given the current limitations of research in this area, further inquiry is necessary to gain a more comprehensive understanding of the impact of generative AI on students' learning experiences and learning outcomes.

In this study, we used a customized chatbot to generate explanations for graphical representations (see **IV.A.3.**). The study is designed as a field study, with students as the target group and the objective of addressing the following key research questions:

How does learning with AI-generated explanations using a custom chatbot affect

RQ1: ... emotional aspects?

RQ2: ... situational interest?

RQ3: ... cognitive load?

RQ4: ... self-efficacy expectations?

RQ5: ... learning performance in a mathematical and physical learning context, particularly in terms of the necessary transfer performance to the physical context?

## IV. METHODS

### A. Study Design

#### 1. Sample

A total of $N = 214$ sixth-grade students (146 female, 66 male, 2 n.a.) at secondary schools in Germany participated voluntarily in the randomized controlled study. The average age of the students was 11.7 years (SD = 0.51).

#### 2. Procedure

The students were randomly assigned to either the experimental (EG) or the control group (CG). Both groups were provided with learning materials on the topic of "proportional relationships" in a mathematical and physical context. The topic had not been covered in class before the study was conducted. The CG was provided with conventional textbook material, comprising a topic overview and explanatory examples. The EG was also provided with the topic overview from the textbook. Instead of the aforementioned examples, the participants were presented with a related explanation of the topic, which was generated by an AI chatbot (see **IV.A.4.**)

---

[2] https://openai.com/index/introducing-gpts/
accessed: 2024-12-14

[3] https://docs.anthropic.com/en/docs/build-with-claude/tool-use
accessed: 2024-12-14

using a previously defined prompt. To eliminate potential bias caused by the use of digital media, both groups engaged in their studies exclusively with paper-based materials. To ensure a fair comparison, the learning time was identical for both groups and amounted to 15 minutes. Moreover, the learners were not informed in advance about the specific type of learning material they would be using.

Prior to the learning phase, demographic data and individual, subject-specific interest in mathematics were collected. This was done in order to ascertain whether there is a discrepancy in mathematical interest between the two groups. Immediately following the learning phase, the dependent variables outlined in section **IV.A.5.** were collected. Participants' emotions and situational interest were recorded. In addition, the intrinsic and extrinsic cognitive load when learning with the materials and the self-efficacy expectation with regard to solving a topic-related task were assessed. Subsequently, in order to measure learning performance and the ability to transfer mathematical content to a physics context, the learners completed a performance test (see **IV.A.6.**) within 30 minutes.

### 3. Chatbot Design

The custom GPT was configured in OpenAI and adapted to the specific learning context and target group. Furthermore, the chatbot specializes in the analysis and interpretation of the various forms of representation for proportional relationships, including graphs, tables, and formulas. It uses language that is both engaging and age-appropriate, making it an appropriate tool for students in the sixth grade of secondary school. The chatbot is capable of providing targeted and individualized support in both learning mathematical content and applying mathematics in physics lessons.

The chatbot was configured in multiple cycles based on the OpenAI GPT-4.0 model. Following each configuration phase, a test phase was conducted to systematically assess whether the chatbot's responses met the predefined criteria. Based on the results of this evaluation, the configuration prompt was successively adjusted after each test phase. In addition, the chatbot was equipped with subject knowledge that corresponds to the core curriculum for the respective grade. In evaluating the quality of the chatbot, particular attention was paid to ensuring that the answers were factually correct and corresponded to the notations used in common textbooks. Additionally, care was taken to ensure that the language used contained motivating vocabulary and was age-appropriate, assuming a learning group between the ages of 10 and 13.

### 4. Material

*a. Schoolbook material.* Both the control and the experimental group received an overview of the topic "Proportional Relationships", taken from a schoolbook [44] for the learning phase. The CG was also provided with additional examples from the same schoolbook that illustrate and explain the topic, thus compensating for the AI-generated explanation of the topic that was provided to the EG.

The textbook material (see supplemental material, chapter **A.a.** and **A.c.** [62]) was selected by a group of experts according to the following criteria, among others:
- A structured overview of the given topic and the forms of presentation (graph, table, formula)
- An overview that does not contain many additional explanations
- The selected examples are consistent, follow on from the content of the overview (chosen from the same book), and provide a sufficiently in-depth presentation of the topic

*b. AI-generated explanation.* In selecting the AI-generated explanation, particular consideration was given to the following criteria:
- The explanation is accurate and complete.
- All three forms of presentation are adequately explained.
- No terms are used that are unfamiliar to the students.
- No discriminatory language is used.

To achieve a satisfactory output, a prompt was initiated, reiterated, and refined until it resulted in the final selected explanation (see supplemental material, chapter **A.b.** [62]). In addition to the prompt, the custom chatbot was presented with an image of the textbook overview.

### 5. Data Collection

*a. Subject-specific interest.* Interest in mathematics as a school subject was measured using five items and, like all subsequent variables, evaluated on a four-point Likert scale. Subject-specific interest was measured before the learning phase and served to examine the comparability of CG and EG. A scale was utilized, based on the one developed by Rakoczy et al. [45]. The scale was shortened from eight to five items for reasons of test economy and the remaining items were adapted to the specific learning context.

*b. Emotions.* To assess the emotional state of the learners during the learning phase, positive-activating (2 items: pleasure, satisfaction) and negative-deactivating emotions (3 items: boredom, frustration, and uncertainty) were measured retrospectively directly after the learning phase. For this purpose, five items were selected and translated into German based on the Achievement Emotions Questionnaire (AEQ) [46–48].

*c. Situational interest.* Situational interest was measured with four items following the students' engagement with the learning materials. In order to maintain a low number of items for the age group, four items were chosen based on a reliable and validated scale by Linnenbrink-Garcia et al. [49], translated into German and adapted to the learning context.

*d. Cognitive load.* To measure the cognitive load during the learning process, a two-factorial model of cognitive load was used in this work, measuring intrinsic cognitive load (ICL) and extrinsic cognitive load (ECL). Although based on a three-factorial model, the 10-item questionnaire developed and validated by Leppink et al. [50] was utilized (Cognitive Load Scale; CLS) to ensure the validity of the measurement. Three items each are used to assess intrinsic cognitive load (ICL) and extrinsic cognitive load (ECL). These six items were translated into German and adapted to the specific context of the learning environment. Furthermore, the original response scale, which ranged from 0 to 10, was reduced to levels 1 to 4 for reasons of consistency with the rest of the questionnaire.

*e. Self-efficacy expectation.* The self-efficacy expectation is evaluated in regard to the independent solving of a topic-related task on the basis of a scale comprising five items. The concept of self-efficacy expectation was first developed by Bandura [51] and the German adaptation was conducted by Schwarzer and Jerusalem [52]. The items utilized in this study are derived from the scale by Jerusalem und Satow [53].

### *6. Performance test*

Following the examination of motivational aspects, the students completed a performance test (see supplemental material, chapter **B.** [62]) that addressed the physical concept of proportional relationships using the "spring scale" as an example. The tasks were developed internally based on standardized textbook tasks. The objective of the performance test is to ascertain whether the students have acquired an understanding of the various ways in which a proportional relationships can be represented and whether they are able to switch between these different forms of representation. In particular, it becomes evident whether the students are capable of transferring mathematical knowledge into the physical context. The test is comprised of four subtasks A, B, C and D, after which students are requested to indicate the degree of certainty associated with their response. They may select one of four options: "completely sure", "rather sure", "unsure", or "guess". In the interest of ensuring the integrity of the evaluation process, students who have not completed a task are instructed to select "guess" to exclude their responses from being considered during the subsequent evaluation.

### B. Data Analysis

All analyses were conducted using R version 4.4.0 [54].

#### *1. Internal consistency*

Cronbach's alpha was calculated for the purpose of evaluating the internal consistency of the individual scales (R package "psych", version 2.4.6.26, function "alpha") [55]. In this context, values with $\alpha > 0.7$ are considered acceptable; values with $\alpha > 0.8$ are considered good and values with $0.9 < \alpha < 1.0$ excellent [56,57].

| Scale | $\alpha =$ | |
|---|---|---|
| Interest in mathematics | 0.92 | excellent |
| Pos.-activ. emotions | 0.71 | acceptable |
| Neg.-deactiv. emotions | 0.76 | acceptable |
| Intr. cognitive load | 0.73 | acceptable |
| Extr. cognitive load | 0.72 | acceptable |
| Situational interest | 0.84 | good |
| Self-efficacy | 0.90 | excellent |

*Table I: Cronbach's alpha indicating scale consistency.*

As the values for Cronbach's alpha are acceptable to excellent and no significant improvements were identified when individual items were removed from the scales, the scales will be retained for further data analysis. In order to proceed with the evaluation, the mean of the recorded values of the items on each scale was calculated for all participants. These resulting scores form the basis for the subsequent data evaluation.

#### *2. Test for normal distribution*

Prior to testing the data of the experimental and control groups for significant deviations, the entire data set is initially evaluated for normal distribution. This is done to enable the drawing of conclusions regarding the applicability of specific statistical tests. Both the data of the EG and those of the CG were tested for normal distribution using the Shapiro-Wilk test with a significance level $\alpha = 0.05$ (R package "stats", function "shapiro.test") [54]. In accordance with the selected significance level, a normal distribution of the data can be assumed if the p-value of the Shapiro-Wilk test is $p > 0.05$. However, this is only the case for the self-efficacy score in the experimental group for the given data set ($p = 0.101$) (see supplemental material, chapter **F.a.** [62]). No variable demonstrated a normal distribution in both groups, therefore non-parametric tests were employed.

### *3. Test for significant deviations*

To test for significant differences between the experimental and control group scores, the Mann-Whitney U test with significance level $\alpha = 0.05$ was used (R package "stats", function "wilcox.test") [54]. This test is standardized for two independent non-normally distributed samples. Cohen's d was calculated to estimate the effect size for significant differences (R package "psych", version 2.4.6.26, function "cohen.d") [55].

### *4. Evaluation of the Performance test*

*a. Interrater reliability.* It was not possible to determine a clear allocation of points for one of the subtasks (task D). The task required the creation of a graph within a coordinate system. A rating system of six categories was devised for the evaluation of the task (see supplemental material, chapter **C.** [62]) and a second, independent rater was consulted. To ascertain the interrater reliability, Cohen's weighted kappa was computed for the two raters and for each category of the rating system (R package "irr", version 0.84.1, function "kappa2") [58]. Quadratic weighting was selected to account for the discrepancy in point allocations across the rating scale (the difference between 0 and 2 carries greater weight than that between 0 and 1). Computing the mean value of the interrater reliability from all categories, results in a mean value of kappa $= 0.8469$ which indicates a very good degree of agreement between the two raters for task D [59–61].[4] To further evaluate the data, the mean of the two ratings is calculated and taken as the score for task D.

*b. Filtering data and calculating scores.* As previously outlined in **IV.A.6.**, the students were asked to evaluate their confidence in their responses following the completion of each subtask. Consequently, the points awarded for solutions marked as "guessed" are set to zero. Furthermore, an item designated as the "overall score" is introduced for the performance test, wherein all subtasks are given equal weighting.

*c. Test for normal distribution.* Prior to the group comparison, the data for the performance test were also evaluated for normal distribution using the Shapiro-Wilk test with $\alpha = 0.05$ (R package "stats", function "shapiro.test") [54]. The results of this analysis indicated that the data did not meet the criteria for a normal distribution.

*d. Test for significant deviations.* Given that the data from the performance test are not normally distributed, the Mann-Whitney U test for independent, non-normally distributed samples is also applied here for the comparative analysis of the groups (R package "stats", function "wilcox.test") [54]. Cohen's d was calculated to estimate the effect size for significant differences (R package "psych", version 2.4.6.26, function "cohen.d") [55].

## V. RESULTS

### A. Mathematics – Grade and Interest

To ensure a variable to assess the comparability of the two groups with regard to performance in math classes, the math grade from the last term's report card was requested.[5] The CG exhibits slightly superior mathematical performance (M = 2.24; SD = 0.88) than the EG (M = 2.38; SD = 0.95), but the Mann-Whitney U test revealed no statistically significant discrepancy between the two groups with respect to their mathematical performance ($p = 0.252 > 0.05$).

### B. Measured variables

#### *1. Subject-specific interest*

The subject-specific interest was evaluated prior to the introduction of the instructional materials. Therefore, the assignment to a group could not have had any effect on the measured results at this point. The data collected show no significant differences between the two groups in terms of interest in mathematics. The

---

[4] A value of 0.60 or higher is considered "good" and a value of 0.8 or higher is considered "very good" or "almost perfect".

[5] In Germany, grades range from 1 to 6, with 1 representing the highest level of achievement and 6 indicating the lowest.

Mann-Whitney U test calculated $p = 0.961$, indicating that there is no statistically significant difference between the two groups.[6]

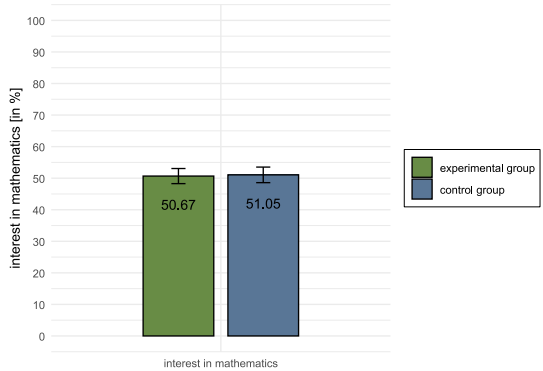

FIG 1: Interest in mathematics with standard error.

### 2. Emotions

The elicited emotions were classified into two categories: positive-activating and negative-deactivating. For the former, a statistically significant difference ($p = 0.00093$, $d = 0.48$, small effect size) was observed in favor of the EG.

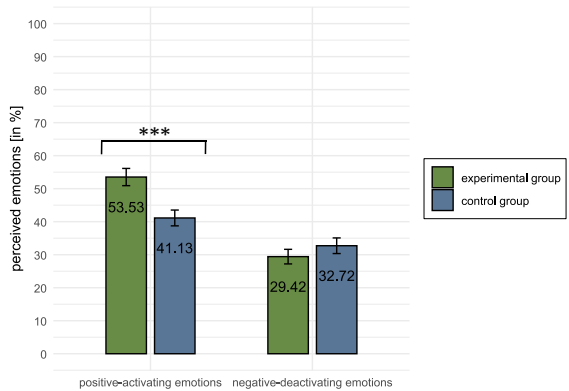

FIG 2: Perceived emotions with standard error and level of significance.

No significant deviation could be detected for negative-deactivating emotions ($p = 0.391$). Only a trend can be identified, which shows that the CG tended to perceive more negative emotions.

### 3. Situational Interest

The measured situational interest was significantly higher in the EG ($p = 0.00223$, $d = 0.45$, small effect size) than in the CG.

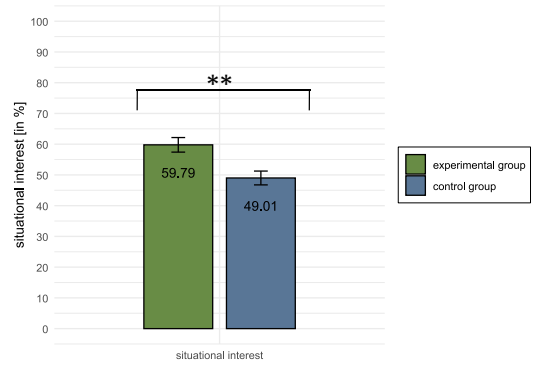

FIG 3: Situational interest with standard error and level of significance.

### 4. Cognitive load

Cognitive load was measured as intrinsic cognitive load (ICL) and extrinsic cognitive load (ECL). In both cases, the CG experienced significantly greater cognitive load than the EG:

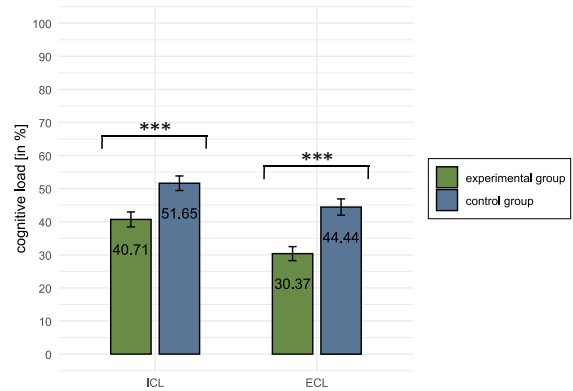

FIG 4: Intrinsic cognitive load (ICL) left and Extrinsic cognitive load (ECL) right, both with standard error and level of significance.

The analysis showed that a significant deviation ($p = 0.00060$, $d = 0.47$, small effect size) was calculated for ICL, and that a significant deviation ($p = 0.0001$, $d = 0.59$, medium effect size) was calculated for ECL.

### 5. Self-efficacy expectation

Self-efficacy was measured in terms of solving a topic-related task. The values of the EG were found to be significantly higher than those of the CG ($p = 0.00001$, $d = 0.63$, medium effect size).

---

[6] The complete results of the Mann-Whitney U test, including those relating to individual items, can be found in the supplemental material, chapter **F.b.** [62].

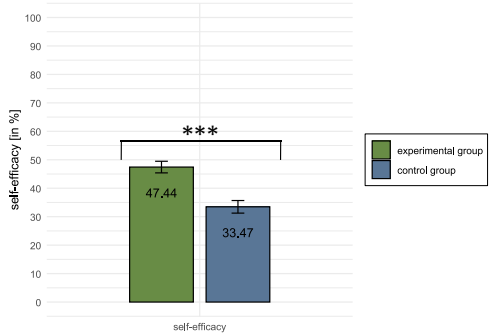

FIG 5: *Self-efficacy with standard error and level of significance.*

The results demonstrate that students who engaged in learning activities with AI-generated explanations exhibited higher levels of self-efficacy than those who utilized textbook materials exclusively.

### C. Performance test

For the overall test score, no significant difference between the two groups could be determined ($p = 0.750$):

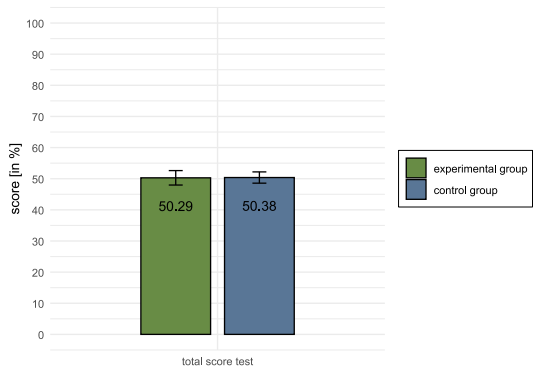

FIG 6: *Overall score of the test with standard error.*

Subsequently, the scores of the individual subtasks were considered. First of all, the assessments of the students' confidence in their responses were compared between the two groups. No significant differences were identified for any of the four subtasks (see supplemental material, chapter **F.b.** [62]). Moreover, the Mann-Whitney U test revealed no statistically significant differences between the EG and CG with regard to subtask A ($p = 0.133$) and D ($p = 0.251$). However, a significant difference was identified for subtask C in which the constant of proportionality of a given proportional relationship should be calculated based on given data.

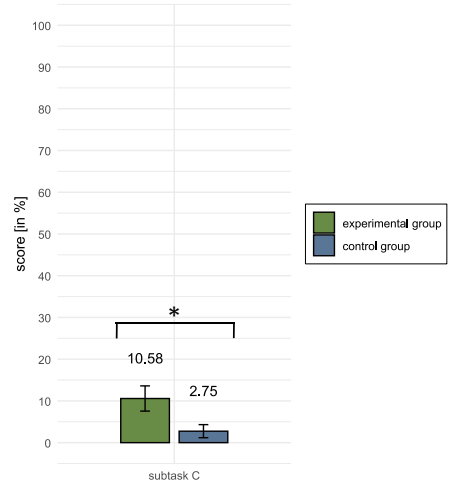

FIG 7: *Score of subtask C with standard error and level of significance.*

The EG achieved a significantly higher score than the CG ($p = 0.022$, $d = 0.32$, small effect size).

The analysis of subtask B also revealed significant differences. In subtask B the students were asked to compute six specific missing values (henceforth, designated as value 1 to value 6) of a proportional relationship, based on one given pair of values, by filling in a corresponding table. For the overall score of subtask B, no significant difference between the two groups could be determined ($p = 0.403$). However, to provide greater insight, the solutions for value 1 to value 6 were also evaluated for notable discrepancies. In this regard, the solutions for value 3 ($p = 0.016$, $d = 0.33$) and value 5 ($p = 0.038$, $d = 0.29$) exhibited significant differences with small effect size:

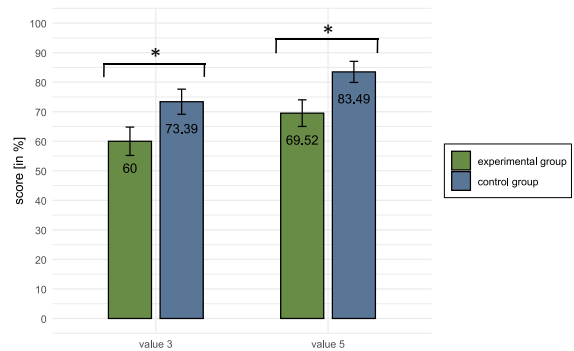

FIG 8: *Score of solutions for values 3 and 5 of subtask B, both with standard error and level of significance.*

In both cases, the students in the CG performed significantly better compared to those in the EG.

### VI. DISCUSSION

As no significant differences were observed between the EG and CG with respect to mathematics grade or subject-specific interest, these two groups were found to be comparable in this regard.

Concerning the influence of emotional aspects (RQ1), the EG exhibited a significantly higher incidence of positive-activating emotions (pleasure and satisfaction) than the CG. This finding is consistent with the results of previous research [29]. With regard to negative-deactivating emotions (boredom, frustration, and uncertainty), only a trend could be identified that suggests that learning with AI-generated explanations may potentially contribute to a reduction of negative emotions. In sum, the results of the evaluation indicate that learning with AI-generated explanations can have a positive impact on emotional states of the learners.

Regarding the influence on situational interest (RQ2), it was found that situational interest was significantly higher in the EG than in the CG. When learning with the explanation generated by the custom chatbot, a higher situational interest could be triggered than when learning with the textbook materials. One potential explanation for this outcome is the nature of the custom chatbot's tone, which is designed to be encouraging and motivating.

The results concerning the effects on cognitive load (RQ3) show significant differences for both intrinsic cognitive load (ICL) and extrinsic cognitive load (ECL) – in both cases the values were higher in the CG. These findings suggest that learning with AI-generated materials reduces cognitive load compared to learning with textbook-only materials. One potential explanation for this outcome is the use of accessible language and a conversational and human-like tone in the chatbot's responses.

Regarding research question RQ4, which concerned self-efficacy expectations, the EG showed significantly higher values than the CG. This indicates that the explanation of the AI chatbot can foster students' confidence in their ability to complete tasks and corresponds to the findings of Canonigo in mathematics [30]. A reason for this may be the encouraging tone of the AI-generated text or the reduced cognitive load, or a combination of both.

The fifth research question RQ5 related to the influence of the learning materials on learning performance in the mathematical and physical learning context, with a particular focus on transfer performance. No clear trends could be identified either in the overall result of the performance test or in the subtasks. For two values of task B (tabular), the CG demonstrated a significantly higher level of performance than the EG. Conversely, in subtask C, the EG performed significantly better. The latter observation may be indicative of a beneficial impact of AI-generated explanations on the transfer of mathematical knowledge into a physics context. This is because the EG demonstrated superior abilities in mathematising the proportional relationship in comparison to the CG. However, the available data is insufficient to permit a definitive conclusion to be drawn. It is not possible to provide a definitive answer to RQ5 based on the data collected in this study. The results indicate a tendency towards both outcomes. A larger-scale study in which students could utilize the chatbot independently and individually would be beneficial to verify the observations and allow for a higher degree of individualization of the feedback. Prior research has demonstrated that the utilization of chatbots can enhance learning outcomes [9,27,30–32].

Whether there is a connection between the learning performance and the previously measured variables cannot be definitively determined on the basis of the present study. Despite the fact that the experimental group exhibited a notable reduction in cognitive load, this did not translate into a statistically significant improvement in performance on the achievement test. It is possible that the lack of independent use of the chatbot in this study may have contributed to the observed outcomes, as it prevented the identification and addressing of individual learning difficulties.

### VII. CONCLUSION AND OUTLOOK

The findings of the present study provide valuable insights into the effects of learning with explanations generated by a custom chatbot on students' learning experience and learning performance. It was shown that learning with these explanations has a significant impact on learners' positive-activating emotions, situational interest, and self-efficacy, while also reducing cognitive load in comparison to traditional learning with textbook materials. Nevertheless, the influence on learning outcomes remains unclear. No notable differences in the overall result of the performance test could be identified.

The generalizability of the results is constrained by different factors: the size of the sample and the brief implementation period (snapshot) as well as the necessity of selecting a particular topic for a specific target group. Additionally, the learning materials were provided in a centralized manner, which resulted in the lack of consideration for individual differences in utilization and the potential for interaction with the chatbot. With regard to learning performance, the data revealed only slight tendencies that suggest a positive influence of the AI-generated explanations on the ability to mathematize. However, due to the limited data basis, no definitive conclusions can be drawn.

Further research could serve to confirm the statements and assumptions posited here through the implementation of larger and longer-term studies with a variety of topics and a wider target group. Moreover, subsequent to the present study, learners should be afforded the opportunity to utilize the AI chatbot independently and individually in order to measure the impact of individualized feedback. To gain a more nuanced understanding of the usability of custom chatbots at the interface between mathematics and physics, it would be prudent to implement a design with a more pronounced emphasis on the task selection and the chatbot design with respect to the transfer of mathematics to physics. Further research with larger samples and a stronger focus on transfer performance would be needed to confirm the results and more fully evaluate the potential of using AI-enhanced learning materials.

Overall, the present work shows that the use of AI in the learning process offers promising possibilities, although further research is needed to fully realize the potential and increase effectiveness in different learning scenarios.

## ACKNOWLEDGMENTS

We would like to express our sincere thanks to all participating schools, teachers and students for contributing to our study.